Dust release from cold ring particles as a mechanism of spoke formation in Saturn's rings

Authors
Naoyuki Hirata[a, *], Hiroshi Kimura[b], Keiji Ohtsuki[a]

Authors' affiliations
[a]Department of Planetology, Kobe University, Rokkodai, Kobe 657-8501, Japan.
[b] Planetary Exploration Research Center (PERC), Chiba Institute of Technology, Narashino, Chiba 275-0016, Japan
* Corresponding Author E-mail address: hirata@tiger.kobe-u.ac.jp

Plain Languages Summary
We noticed that the spokes are very similar to a mechanism known as the Lunar Horizon Glow or Dust levitation. Base on the formation mechanism proposed by the Lunar Horizon Glow, we developed a novel model for the spoke formation. As a result, we found that our model explains observational features of the spokes including their longitudinal location, lifetime, radial expansion velocity, and seasonality. We predict that ring particles have intense electric fields across their terminator that cause dust release. The gas adsorption, such as $O_2$ ring atmosphere reduces the surface energy of ring particles, which would explain the seasonal appearance of the spokes. Previous models proposed a plasma environment, dense plasma induced by impacts of meteorites, field-aligned electron beams flowing from the magnetotail into the auroral regions of Saturn, or electron beams induced by lightning in Saturn's atmosphere, to create the spokes, however, our model indicates that those models are unnecessary.

# Highlights

- We propose a novel model for the formation of spokes in Saturn's main rings
- There are two key processes in dust release from ring particles
- One is intense electric fields across the terminator of ring particles
- The other is adsorption of an $O_2$ ring atmosphere onto the surface of ring particles
- Our model explains lifetime and seasonality of the spokes.

# Abstract

Spokes in Saturn's rings are radially-extended structures consisting of dust grains. Although spacecraft and space telescope observations have revealed various detailed features of the spokes and their time variation, their formation mechanism is still under debate. Previous models examined charging mechanisms to attempt at explaining dust release from cm-sized ring particles; however, the attempt has been unsuccessful, because the electrostatic force caused by such charging mechanisms is much weaker than the cohesive force acting on dust grains at ordinary conditions in the ring environment. Here we propose a novel model for the formation of the spokes, where the temperature dependence of cohesion plays an essential role. Ring particles with a temperature below 60K adsorb an $O_2$ ring atmosphere, which facilitates release of dust grains from them by a reduction in the cohesive force between the grains and the particles on the morning ansa. Then, intense electrostatic forces sufficient to overcome the cohesive force are generated on the surface of ring particles and the released dust grains form the structure of spokes. Our model explains observational features of the spokes including their longitudinal location, lifetime, radial expansion velocity, and seasonality.

# 1. Introduction

Spokes in Saturn's rings were observed in detail by the two Voyager spacecraft, the Hubble Space Telescope, and the Cassini spacecraft (e.g., Horányi et al. 2009). They are seen only in the outer B-ring, the optically thickest region of the main rings (Grün et al. 1983, Smith et al. 1981). The motion of the spokes is close to the purely Keplerian motion but is affected by the corotation of Saturn's magnetic field (Grün et al. 1983, Smith et al. 1981, Eplee and Smith 1984). Some of the spokes are affected by the corotation significantly and show an intermediate motion between Keplerian and the corotation (Mitchell et al. 2013). The spokes, when they appear, expand radially inward and outward with a speed of about 0.5 - 1 km/s (e.g. Mitchell et al. 2013). The spokes are observed as dark markings at low phase angles, while as bright markings at high phase angles, which is attributed to forward scattering of sunlight by micron-sized particles, thus shows that the spokes are composed of dust grains levitated above the ring plane (Smith et al. 1981, Doyle and Grün 1990, McGhee et al. 2005, D'Aversa et al. 2010). The appearance of the spokes is predominant on the morning ansa (the part of the ring just emerged from Saturn's shadow) and they typically disappear in less than several hours, which is about a half of the orbital period (Porco and Danielson 1982, Smith et al. 1981, Grün et al. 1983). Moreover, the spoke formation is a seasonal phenomenon related to the ring opening angle relative to the Sun (McGhee et al. 2005, Mitchell et al. 2006); the spokes appear when Saturn is near its equinox, whereas disappear when it is near its solstice and the tilt angle is approximately larger than 15 degree.

Previous models for the formation of the spokes mostly focused on the charging mechanism of ring particles; it is argued that dense plasma induced by impacts of meteorites onto the rings charges the particles negatively, and the electrostatic repulsive force acting on dust grains overcomes the gravity, resulting in the dust release (Goertz and Morfill 1983); instead of impact-induced plasma, it is also proposed that field-aligned electron beams flowing from the magnetotail into the auroral regions of Saturn or electron beams induced by lightning in Saturn's atmosphere charge up ring particles (Jones et al. 2006, Horanyi, Morfill and Cravens 2010, Hill and Mendis 1981). Previous models proposed that the seasonal appearance of the spokes is attributed to their immediate destruction under the plasma environment of the ring near the solstice (Mitchell et al. 2006, Farrell et al. 2006). However, these models overlooked the cohesive force (i.e., the van der Waals force to a large extent) between μm-sized dust grains and meter-sized ring particles, which in fact plays an essential role.

In the present work, we propose a novel model for the formation of the spokes, where the temperature dependence of cohesion plays an essential role. The structure of this paper is as follows. In Section 2, we describe cohesive force and electrostatic force acting on dust grains on the surface of ring particles and demonstrate that previously-proposed charging mechanisms cannot overcome the cohesive forces. Furthermore, we demonstrate that the combination of intense electric fields across the terminator of ring particles (Section 3) and reduction of cohesion due to adsorption of $O_2$ gas molecules (Section 4) allows dust release from ring particles, and consequently, the released dust particles are observed as the spokes in the ring plane (Section 5). Our model basically follows models for the formation of the lunar horizon glow, caused by electrostatically levitated dust particles above the lunar terminator. In Section 6, we demonstrate that our model explains observational features of the spokes including their longitudinal location, lifetime, radial expansion velocity, and seasonality.

## 2. Cohesive force and electrostatic force acting on a dust grain
### 2.1. Cohesive force

The cohesive force acting on dust grains with planetary surfaces is much greater than the gravitational force (e.g., Scheeres et al. 2010, Hartzell and Scheeres 2011). Even if µm-sized dust grains are floating in Saturn's rings, they would accumulate on cm-sized ring particles immediately and could be hardly released from the particles owing to the cohesion (Bodrova et al. 2012, Ohtsuki et al. 2020) (see also Appendix A). The cohesive force between two identical dust grains is given by (Johnson et al. 1971)

$F_{co} = \frac{3}{2}\pi\gamma a$ , (1)

where $\gamma$ is the surface energy and $a$ is the radius of the grain. On the other hand, the gravitational force acting on a dust grain on an airless body is given by

$F_g = \frac{4}{3}\pi a^3 \rho g$ , (2)

where $\rho$ is the density of the grain and $g$ is the surface gravity of the airless body. The equations are assuming spherical particles. Assuming an ice grain with $\gamma$ = 0.19 Jm$^{-2}$ (e.g., Gundlach et al. 2011), $\rho$ = 900 kgm$^{-3}$, and $a$ = 1 µm on the surface of a ring particle with a radius of 1 m ($g$ = 2.5×10$^{-7}$ ms$^{-2}$), the cohesive force acting on the grain is 10$^{15}$ times greater than the gravity. Interestingly, the spokes consist of dust grains floating above the ring plane, despite such a strong cohesive force. This indicates that the detachment of the grains sticking to ring particles is a key process for the formation of the spokes.

### 2.2. Electrostatic force

The physics of the electrostatic force acting on dust particles depends on the conductivity of the particles (Figure 1) (Jones 1995, Sow et al. 2013, Kok and Renno 2006, von Holstein-Rathlou et al. 2012, Nakajima and Matsuyama 2002). For conducting particles, the charge in an electric field is enhanced by the electrostatic induction. For example, a particle placed in an electric field perpendicular to a positively charged surface would polarize its top to be positive and its bottom to be negative. If its bottom is in contact with other conducting particles, electrons will flow to other particles. Consequently, particles on the top of the surface are positively charged and are able to lift off from the surface. In contrast, for an insulating particle, the charge in a parallel electric field is not enhanced. Insulating particles would be attracted with each other owing to dielectric polarization under intense electric fields.

For a conducting particle, the electrostatic forces acting on the particle in a parallel electric field ($E$) can be described by (van den Bosch et al. 1995, Lebedev and Skalskaya 1962, Arp and Mason 1977, Kimura et al. 2014)

$$F_e = \frac{6}{\pi^2}\left(\zeta(3) + \frac{1}{6}\right) Q_m^s E \ , (3)$$

where $\zeta$ is Riemann's zeta function ($\zeta(3)$ = 1.202) and $Q_m^s$ is the maximum induction charge of dust grains with

$$Q_m^s = \max\left\{e \ , \frac{2}{3}\pi^3 a^2 \varepsilon_0 E f_r\right\} \ , (4)$$

where $e$ is the elementary charge, $\varepsilon_0$ is the vacuum permittivity, $a$ is the radius of the grain, and $f_r$ is the roughness factor. We use $f_r$ = 10, which is the average value of lunar soil (Cadenhead et al. 1977). As a result, we obtain

$$F_e = 4\pi \left(\zeta(3) + \frac{1}{6}\right) \varepsilon_0 f_r E^2 a^2 \ . (5)$$

For an insulating particle, the repulsive force acting on the particle in a parallel electric field ($E$) can be described by (Lebedev and Skalskaya 1962)

$$F_e = -\frac{Q^2}{16\pi\varepsilon_0 a^2} + \left(1 + \frac{\delta}{2}\right) QE - \frac{3}{2}\delta^2 \pi \varepsilon_0 a^2 E^2 \ , (6)$$

where $Q$ is the charge of the grain, and $\delta$ is a constant described by

$$\delta = \frac{\varepsilon_p - 1}{\varepsilon_p + 2} \ , (7)$$

where $\varepsilon_p$ is the static dielectric constant. The dielectric constant for ice is greater than 100 (Johari and Whalley 1981) and therefore we can assume $\delta \sim 1$. On the other hand, if icy ring particles are covered with $O_2$ molecules, the dielectric constant is $\varepsilon_p$=2.46 ($\delta \sim$ 0.33) (Pilla et al. 2008). Note that it is probable that icy ring particles are covered with $O_2$ molecules when the Saturn is near the equinox (see also Section 4). The charges on a dusty surface owing to solar irradiation or plasma can be assumed to be uniformly distributed over the surface of each grain, and we can therefore assume the charge of

each dust grain is given by

$$Q = \max\{e, 4\pi\varepsilon_0 a\Phi\} \quad , (8)$$

where $\Phi$ is the surface potential of each grain. Note that the surface potential in a steady state during equinox is simply determined by the electron energy ($kT_e$) of the plasma (Spitzer Jr. 2008):

$$\Phi = -2.5kT_e \quad . (9)$$

The maximum value of Eq. (6) can be obtained when

$$Q = \pm 8\left(1 + \frac{\delta}{2}\right)\pi\varepsilon_0 a^2 E \quad (10)$$

as

$$F_e = \frac{\pi}{2}(-\delta^2 + 8\delta + 8)\varepsilon_0 E^2 a^2 \quad . (11)$$

Although the electrical conductivity of grains in Saturn's rings is poorly known, $O_2$ ions may assist in elevating the surface conductivity of icy grains by proton jumping (Appendix B).

### 2.3. Electrostatic force induced by previously-proposed charging mechanisms

We demonstrate that previously-proposed charging mechanisms for the spoke formation cannot overcome the cohesive forces and the mechanisms are insufficient to cause dust release from ring particles. Generally, the charges of the surface of a spherical body induced by plasma absorption are considered to be distributed uniformly over the surface (Flanagan and Goree 2006). Assuming a uniformly distributed charge on a dielectric sphere with a radius of $r$, the total charge of the sphere ($Q$) and the electric potential on the surface of the sphere ($\Phi$) have a relation:

$$\Phi = \frac{Q}{4\pi\varepsilon_0 r} \quad , (12)$$

where $\varepsilon_0$ is the permittivity of free space (Flanagan and Goree 2006). The strength of the electric fields ($E$) is given by (Flanagan and Goree 2006)

$$E = \frac{Q}{4\pi\varepsilon_0 r^2}. (13)$$

As a result, we obtain the relation:

$$E = \frac{\Phi}{r} \quad . (14)$$

The electrostatic forces acting on a dust grain is determined by this equation and Eq. (5) or Eq. (11).

The electric surface potential of a spherical particle is determined by the energy and density of ions, electrons, secondary electrons, or photoelectrons, but it approximately corresponds to the electron temperature (Manka 1973). As an example, if we assume the typical electron temperature at the B ring, or the order of 1 eV (Farrell et al. 2017), the surface potential of Saturn's ring particles can be estimated to be roughly on the order of -1V. Then, the strength of electric fields on a ring particle with $r$=1m are the order of 1V/m. Assuming dust grains with a radius of $a$ = 10µm and surface energy of $\gamma$= 0.190 J·m$^{-2}$ on the ring particle with $r$=1m, we find that the electrostatic force obtained from electric fields of 1V/m and Eq. (5) is by a factor of $10^{14}$ ~ $10^{15}$ weaker than the cohesive force shown in Eq. (1). Comparing between Eq. (1) with (5) or (11), the electrostatic force is able to overcome the cohesive force only if the surface has an electric potential greater than $10^7$ V. This huge potential requires unusually high electron temperature, $10^7$ eV, however previously-proposed charging models cannot provide such high temperature; (1) the electron temperature of impact-induced plasma proposed by Goertz and Morfill (1983) is 2 eV, (2) the electron temperature of field-aligned electron beams flowing from the magnetotail into the auroral regions proposed by Hill and Mendis (1981) is on the order of $10^3$ eV, and (3) the electron temperature of the lightning-induced electron beams proposed by Jones et al. (2006) and Horanyi et al. (2010) is on the order of $10^5$ eV. Even in the case of the lightning-induced electron beams, the electrostatic force is weaker by a factor of $10^4$ than the cohesive force. In addition, electric potential exceeding the order of 1000 V become unstable in vacuum because the electric discharge in vacuum takes place at 1000 V. Consequently, none of previously-proposed charging mechanisms alone can explain the detachment of spoke materials from the ring particles against the cohesive forces.

3. Generation of an intense electric field on particle surfaces in our model

Instead of uniformly-distributed electric potential induced by previously-proposed charging mechanisms, here we propose an intense electric field generated across the terminator of ring particles (Figure 2). It is proposed that dust grains on the Moon are electrically lofted above the surface and observed as the lunar horizon glow (e.g., Colwell et al. 2007) (see also Appendix C). The conventionally accepted models consider dust release by intense electric fields generated across the lunar terminator (i.e., the line separating the sunlit and shadowed sides) (Criswell and De 1977). Photoelectron emission owing to solar UV irradiation charges up the lit surface of the Moon positively, and the absorption of solar wind electrons at the unlit side and/or photoelectrons emitted from the sunlit side charge up negatively (Manka 1973, De and Criswell 1977). These two areas are adjacent to each other at the terminator and produce intense electric fields (Criswell 1973, Lee 1996). On the lunar terminator, it is proposed that a potential difference of 1000 V is generated across a cm-scale boundary between the lit and unlit areas, which generates the electric field strength of ~ $10^5$ V/m (this is obtained by dividing the potential difference, 1000 V, by the distance, 1cm) (Criswell 1973, Lee 1996). This is a rough estimate; for example, the electric field strength across a 10 cm-scale boundary (i.e., an area 10 cm away from the boundary) would be $10^4$ V/m, while the electric field strength across a mm-scale boundary (i.e., an area 1 mm away from the boundary) would be $10^6$ V/m. Nonetheless, in the former case, although such electric field strength affects a wider range of surface area and more dust grains, the strength is too weak to detach dust grains against cohesion. In the latter case, although such electric field strength is very strong, it affects a narrower range of surface area and less dust grains. De and Criswell (1977) and Lee (1996) used the value of $3.0×10^5$ V/m as the electric field strength across lunar terminator. As on the Moon, it is observed that Hyperion has intense electric fields at its terminator (Nordheim et al. 2014). An experimental study, setting a plane with a potential of 10 V adjacent to a plane with a potential of 0 V owing to differential UV charging, showed that strong electric fields, ~1 kV/m, can be generated across the boundary (Wang et al. 2007).

Similar electric fields can be expected for the terminators of ring particles in the Saturn system (Figure 2), for the following reasons; (1) Since the potential difference across the terminator is constrained only by the highest-energy of photoelectrons (500 ~ 1500 eV) ejected from the surface (Lee 1996, De and Criswell 1977), it does not depend on the distance from the Sun; (2) Neutralizing effects due to plasma adsorption onto ring particles, which are able to inhibit the generation of intense electric fields across the terminator, are negligible at the B ring, since the electron density and temperature at the B ring are significantly low, 0.04 - 0.4 electron·cm$^{-3}$ and 1 eV (Farrell et al. 2017). These values are much smaller than those at the Moon, 5 electron·cm$^{-3}$ and 10 eV (Mendis et al. 1981, Colwell et al. 2005) or Hyperion, 0.05 electron·cm$^{-3}$, and 60 eV (Nordheim et al. 2014), and this indicates that ring particles in the B ring can generate intense electric fields more easily than Hyperion or the Moon. Throughout this paper, we basically adopt the value of $3.0×10^5$ V/m, following De and Criswell (1977) and Lee (1996); however, we note that electric fields of $3.0×10^4$ V/m or $3.0×10^6$ V/m would also exist across the terminator. This electric field across the terminator does not require any particular charging mechanism as proposed by the previous models. On the one hand, the so-called supercharging at the progression of sunset is expected to increase the electric field in the lunar terminator region (Criswell and De 1977), but this effect is most likely negligible for ring particles owing to their fast spins. On the other hand, the aggregate structure of dust grains could enhance the charge-to-mass ratio of the aggregates (Kimura et al. 2015). Any charging mechanism of elevating local electric fields and the charge-to-mass ratio might assist our spoke formation mechanism, but these are not mutually exclusive models.

We note that the so-called charge patch model (e.g. Wang et al. 2016) has been recently proposed as an alternative dust release mechanism based on laboratory experiments. However, it is likely that cohesive force between dust grains was significantly reduced in their experiment setup compared to ultra-high-vacuum conditions in space, where the cohesion between dust grains is expected to be much stronger (Appendix D). Therefore, we think that reduction of cohesive force is essential for the release of dust grains as we describe in detail in the next section.

## 4. Reduction of cohesion on particle surfaces

On the Moon, dust grains are covered with water molecules such as OH or $H_2O$ (Hibbitts et al. 2011, Pieters et al. 2009, Clark 2009), which play a key role in reducing the cohesive force acting on the grains significantly for the dust release (Perko et al. 2001). The surface energy of dehydroxylated amorphous silica in a vacuum (not covered with molecules) is $\gamma = 0.25$ J m$^{-2}$ (Kimura et al. 2015). If dust grains on the Moon have this strength of surface energy, the lunar horizon glow does not occur owing to intense cohesion. Then, adsorption of molecules such as OH or $H_2O$ onto dust grains plays a key role in reducing the surface energy sufficiently to detach dust grains from the lunar surface. For example, Perko et al. (2001) suggest that the cohesive force acting on lunar dust grains becomes weaker during the lunar nighttime due to adsorption of molecules prompted by the lower temperature. Hibbitts et al. (2011) utilize lunar analog materials and show experimentally that $H_2O$ molecules can be accumulated and chemisorbed onto soil especially in the mornings and evenings of the Moon. The gas adsorption reduces the surface energy of a grain down to ~0.01 times of the dehydroxylated amorphous silica in a vacuum, as summarized in Fig. 1 of Kimura et al. (2015). In fact, silica in air (covered with layers of $H_2O$ molecules) has the surface energy of $\gamma = 0.025$ J m$^{-2}$ (Kendall et al. 1987), and silica in dry air (with a lesser amount of $H_2O$ molecules) has $\gamma = 0.006$ J m$^{-2}$ (Vigil et al. 1994). The very low surface energy in dry air results from the small size of the capillary neck formed by a low amount of $H_2O$ molecules in the progress of evaporation (Vigil et al. 1994). The surface energy of grains is again described by $\gamma(1-\phi_r)$, where $\phi_r$ is the reduction factor due to the roughness of grain surfaces ($0 < \phi_r < 1$) (Kimura et al. 2014). The quantity $(1-\phi_r)$ is defined as the ratio of the cohesive force for rough surfaces to that for smooth surfaces. This parameter describes an effect of irregularity or roughness of grains (including a wide range of particle shapes, angularity, orientation and the number of contacts). The reduction factor due to adsorption of gas molecules then takes between $(1-\phi_r) = 1 \sim 0.01$ as an empirical value.

In the case of Saturn's rings, a tenuous $O_2$ atmosphere at the main ring was observed by the Cassini spacecraft. When the $O_2$ molecules are adsorbed onto the surface of dust grains, we would expect reduction in cohesion in a similar way to the lunar horizon glow. The column density of $O_2$ molecules per 1 $cm^2$ of ring plane is on the order of $10^{11}$ molecular $cm^{-2}$ near the equinox and on the order of $10^{12}$ molecular $cm^{-2}$ near the solstice, which indicate that the amount of $O_2$ gas molecules on ring particles is monolayer (Appendix E). Because near-infrared spectra of the B ring are dominated by strong adsorption features due to crystalline water ice (Poulet et al. 2003), we estimate the surface energy of the ring particles to be that of crystalline water ice, $\gamma_{H_2O} = 0.19$ J·$m^{-2}$ (Gundlach et al. 2011) when grains are not covered with any gas molecules. The surface energy of solid $O_2$ is $\gamma_{O_2} = 3.8 \times 10^{-2}$ J·$m^{-2}$ (Lemmon and Penoncello 1994). The surface energy of icy grains covered with $O_2$ can be estimated from the relation (Israelachvili 2011):

$$\gamma = \left(\sqrt{\gamma_{H_2O}} - \sqrt{\gamma_{O_2}}\right)^2. \quad (15)$$

Consequently, we obtain $\gamma = 5.8 \times 10^{-2}$ J·$m^{-2}$ for $H_2O$ ice surfaces with a layer of $O_2$. In addition, in the progress of $O_2$ sublimation, the empirical roughness factor would be $(1-\phi_r) \sim 0.01$. Therefore, we can assume $\gamma(1-\phi_r) = 5.8 \times 10^{-4}$ J·$m^{-2}$ as a value of the surface energy of ring particles in the progress of the $O_2$ sublimation.

An $O_2$ molecule, which is produced by UV photosputtering of $H_2O$ ice, subsequent photoionization of $O_2$, and UV photodissociation of $H_2O$ vapor, collides with icy ring particles about 1000 times during its lifetime (e.g., Tseng et al. 2010, Johnson et al. 2006). The sticking coefficient of tenuous $O_2$ gas molecules onto nonporous ice grains is negligible above 70 K, dramatically increases at 60 K, and reaches almost complete adsorption below 55 K (He et al. 2016). Therefore, the surface energy of ring particles changes drastically for the temperature of ring particles below 60 K and above 60 K.

Ring particles are dominantly made of water ice, but likely include very small amount of organic materials (so-called tholin), nanophase hydrated iron oxides, carbon, silicates, crystalline hematite, metallic iron, and troilite (Ciarniello et al. 2019). Although there is no study about the surface energy of ice grains with such contaminants, it is unlikely that these minor components affect the surface energy of icy ring particles because the amount of such minor components is roughly one ten-thousandth (i.e., the surfaces of particles of the B ring is dominantly exposed to ice molecules). A large amount of $O^+$ in the ring atmosphere (e.g. Tseng et al. 2010, Johnson et al. 2006) would also have little influence, because it is not adsorbed on the surface of ring particles at the temperature of Saturn's ring.

## 5. Results

We calculated the electrostatic forces acting on a dust grain on the ring particle as shown in above, and compared the cohesive and electrostatic forces acting on dust grains as a function of grain size (Figure 3). We find that (1) the electrostatic force acting on μm-sized dust grains never overcomes the cohesive force acting on the dust grains unless the surface energy of the dust grains is reduced by three orders of magnitude and (2) the electrostatic force overcomes the cohesive force if $O_2$ molecules are adsorbed on the surface of dust grain.

In our model, the emergence of spokes in the B ring inevitably depends on the solar elevation angle and shows a seasonable variation. Then, the night-side temperature of ring particles should be the key parameter in the sequence of the dust release, because (1) dust release occurs at the terminator of ring particles (i.e. dawn or dusk), (2) $O_2$ molecular gas preferentially adsorbs onto the night-side of ring particles owing to its lower temperature, and (3) at the moment of dawn (owing to the rotation of ring particles or egress from the Saturn's shadow), the surface energy of dust grains are reduced in the progress of $O_2$ sublimation. Thermal modeling based on Cassini observations show that the temperature of the unlit side of the B ring becomes lower than 60K when $B < 15°$ (Ferrari and Reffet 2013, Morishima et al. 2016). In general, a particle frequently moves between the sunlit and unlit faces even in the dense B ring (e.g. Morishima et al. 2010). The particle gets heated up when it is near the sunlit face, and it gets cooled down when in the unlit face. It is not unusual that well-cooled ring particles are in the sunlit side. We can consider that the condition for spokes to occur is satisfied when the ring opening angle is less than 15 degrees. Figure 4A shows the seasonal variation of the ring opening angle and the seasonal appearance of the spokes based on the previous observations. Figure 4B shows the seasonal variation of the B-ring temperature based on the thermal modeling and the prediction of spoke appearance based on our model. This clearly indicates that our model prediction agrees well with the seasonality of the observed spokes. On the other hand, near the equinox with $B < 3°$, the temperature of the A ring falls below 60K (Morishima et al. 2016). Therefore, dust release from ring particles can be expected also in the A ring near the equinox, but spokes have not been discovered in the A ring so far. This may be due to insufficient amount of dust for the generation of spokes and/or extremely poor visibility of the generated spokes in the region (Appendix F).

## 6. Discussion

Our model can also explain many of other observed features of the spokes. The ring particles are cooled down while they pass through Saturn's shadow (for the period of ~2 hours) and heated up immediately after they emerge from the shadow. Because the surface energy of the grains should be the smallest when sublimation of $O_2$ molecules is in progress, dust release is expected to take place more easily at the morning ansa of the ring plane, compared with the evening ansa. This accounts for the observation that the spokes predominantly appear on the morning ansa.

According to Lee (1996) and De and Criswell (1977), the timescale to reach a potential of 1000V at the terminator is about $10^2$, $10^3$, and $10^4$ seconds for a saturnian ring particle with a radius of 10, 1, and 0.1 m, respectively. Note that particles in the B ring likely have a size distribution from 0.3 to 20 m (French and Nicholson 2000). Therefore, our model predicts that there is a time lag of approximately $10^4$ seconds (3 hours) for the beginning of dust release from the smallest particles after initial dust release from the largest particles and the amount of released dust grains reaches the maximum at about 3 hours after initial dust release. At the moment of dust release, dust grains on ring particles are charged up as shown in Eq. (4) or (10). Therefore, their motion after released is affected by Saturn's magnetic field, which is consistent with the observed non-Keplerian motion of the spokes.

After the release, the grains will accumulate onto the ring particles that have become $O_2$-free and recovered the strong cohesive force. Although the acceleration of a dust grain at launch involves complicated physical processes and thus it is difficult to estimate its launch velocity, the vertical launch velocity of dust grains from the lunar surface in the case of the lunar horizon glow is observationally determined to be ~1 m/s (Rennilson and Criswell 1974). This launch velocity is consistent with Wang et al. (2016), who have experimentally obtained 0.6 m/s as the vertical launch velocity of dust grains. Assuming that a launch velocity of a dust grain on the ring particles is also ~1 m/s, the dust grain is able to stick adhesively to ring particles at first collision, because Eq. (A1) in Appendix A shows that the dust grain is able to stick adhesively onto ring particles at collision when the relative velocity upon collision is smaller than ~1 m/s. The mean free time of a freely floating dust grain in the ring can be estimated from the collision frequency ($\omega_c$) of a ring particle in Saturn's rings. It is

$$\omega_c \approx 3\Omega\tau, \quad (16)$$

where $\Omega$ is the Kepler frequency and $\tau$ is the dynamical optical depth, based on Eq. 14.2 in Schmidt et al. (2009). This is the expression for macroscopic particles in a Keplerian motion, but is applicable to micron-sized particles unless they receive forces other than Saturn's gravity. The Kepler frequency is obtained by

$$\Omega = \sqrt{\frac{GM}{r^3}}, \quad (17)$$

where $M$ is the mass of Saturn and $r$ is the distance from Saturn. Because the mean collision time of a freely floating dust grain ($T_c$) is the reciprocal of Eq. (16), we obtain

$$T_c \approx \frac{1}{3\Omega\tau}. \quad (18)$$

Assuming $r$ = 100000 km, $M$ = 5.68 × 10$^{26}$ kg, and $\tau$ = 4 (e.g., Schmidt et al. 2009), we obtain $T_c$ = 1 hours. In other words, it takes approximately 4 hour from the beginning of dust release until these dust grains re-accumulate onto ring particles. This agrees well with the typical lifetime of the spokes.

In our model, dust release simultaneously begins at any radial distance. However, due to the radial gradient in the spatial density of ring particles (i.e. radial variation of the optical depth of the main ring plane), there should be a time lag in its visibility along the radial direction. We expect that in the densest part of the B ring, the optical depth of dust grains released from large ring particles alone is already high enough for the detection of spokes, while in optically thin parts of the B ring, not only dust grains released from the largest ring particles, but also those from smaller ring particles are required for the optical depth of spokes to reach a critical value for the spoke to be detected by currently available instruments. Because the radial length of the optically thickest part of B ring is roughly 5000 – 10000 km and the time lag in the visibility of spokes is at most on the order of 10$^4$ seconds, the radial expansion velocity would be up to 0.5 - 1 km/s. This velocity coincides with the radial expansion velocity of spokes derived from optical images of spoke observations.

Our model predicts that the spokes should not appear inside of Saturn's shadow because ultraviolet irradiation is essential, and that the spoke materials should be released from the terminator of cm-sized to meter-sized ring particles. The planes of the terminators of ring particles near the equinox become nearly perpendicular to the ring plane, and therefore, the electric fields on average also become nearly perpendicular to the ring plane, which would help the spoke materials to move above the ring plane. While the detachment of dust grains from the surface of ring particles alone cannot explain the intermittent appearance of the spokes, the spoke formation may involve a self-destructive process in which electrostatic lofting of dust grains poses a major threat to other grains lagging behind in electrostatic lofting (Appendix G). Ring particles at the equinox are barely illuminated by solar UV radiation, and therefore, near equinox, the UV photons reflected by Saturn would play a major role in the charging mechanism of the particles (Appendix H). Future observations as well as further analysis of existing data could test our model based on this view.

It is often argued that the main rings are deficient in µm-sized dust grains. Our view indicates that there are much more abundant small dust grains than have been thought and that most of them stick to ring particles by the Van der Waals force. When such grains are released, they are observed as the spokes. This would be important for the lifetime of Saturn's rings, because mechanisms to erode the ring particles highly depend on the particle size. For example, although micro-meteoroid weathering (Ip and Mendis 1983) or photodissociation (Ip 1995) has been proposed as the dominant mechanism to erode ring particles, mechanisms preferentially acting on dust grains, such as sublimation, sputtering, or the Poynting–Robertson effect, might be important alternative candidates for the erosion of the rings. Recent studies of the so-called ring rain (i.e., transport of tiny ring grains along the magnetic field lines from the ring to Saturn's atmosphere) proposed by O'Donoghue et al. (2019) and Hsu et al. (2018) also seem to support this view. More detailed analysis of observed data to constrain the amount of dust grains in the rings, including those observed as spokes, would provide new insights into the evolution and the fate of the rings.

## 7. Conclusion

We proposed a novel model for the formation of the spokes. Here we summarize our model (see also Fig.2). Solar UV irradiation charges up the dayside of a saturnian ring particle positively due to photoelectron emission, while sticking of emitted photoelectrons or background plasma electrons charge up the nightside of the ring particle negatively. On the terminator of the particle, positively charged areas and negatively charged areas are adjacent to each other, which generate intense electric fields across the terminator. The electric potential deference across the terminator reaches the order of 1000 V, which causes intense electric fields on the order of $3.0\times10^5$ V/m. In addition, adsorption of $O_2$ gas molecules in the ring atmosphere onto the surface of ring particles plays a role in significant reduction of cohesion between ring particles, when the temperature of the nightside of the particle falls below 60K. The combination of intense electric fields across the terminator of ring particles and reduction of cohesion due to adsorption of $O_2$ gas molecules allows dust release, and consequently, the released dust particles are observed as the spokes in the ring plane. This model is basically similar to models for the formation of the lunar horizon glow. Our model explains many of observational features of the spokes including their longitudinal location, lifetime, radial expansion velocity, and seasonality.

## Acknowledgments

The authors would like to thank Shinsuke Takasao and Wing-Huen Ip for useful discussion on electric currents and plasma motion in the Saturn system. We appreciate two anonymous reviewers for their comments. This work was partly supported by JSPS Grants-in-Aid for Scientific Research Nos. 23244027, 26400230, 15K05273, 19H05085 (H. K.), 20K14538, 20H04614 (N. H.), 15H03716, 18K11334, and 21H00043 (K.O).

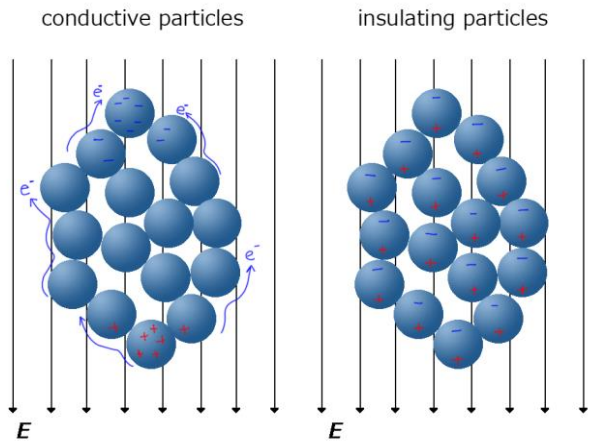

**Figure 1.** Explanations of the electrostatic induction (left) and the dielectric polarization (right). (left) Suppose that an aggregate consisting of conducting particles is placed in a parallel electric field. Electrons move through the conductive aggregate by the electric fields. Positive or negative charges on particles on the top or bottom of the aggregate are enhanced significantly, and the great electric forces act on the particles. This makes the dust release easier. (right) Suppose that an aggregate consisting of insulating particles is placed in a parallel electric field. Because of the dielectric polarizations of each particle, the attractive forces act on each other. This makes the dust release harder.

**Figure 2.** Concept of our model. (1, 2) Solar UV irradiation charges up the dayside of a meter-sized ring particle positively due to photoelectron emission. The Saturnshine could be the dominant ultraviolet source for ring particles near the equinox. (3) At the same time, emitted photoelectron or background plasma absorptions charge up the nightside of the ring particle negatively. (4) On the terminator of the particle, positively charged areas and negatively charged areas are adjacent to each other, which generate intense electric fields across the terminator. (5) When the temperature of the nightside of the particle falls below 60K (i.e. near equinox), the surface energy of the particle is reduced sufficiently to allow dust release. (6) Consequently, the dust release takes place.

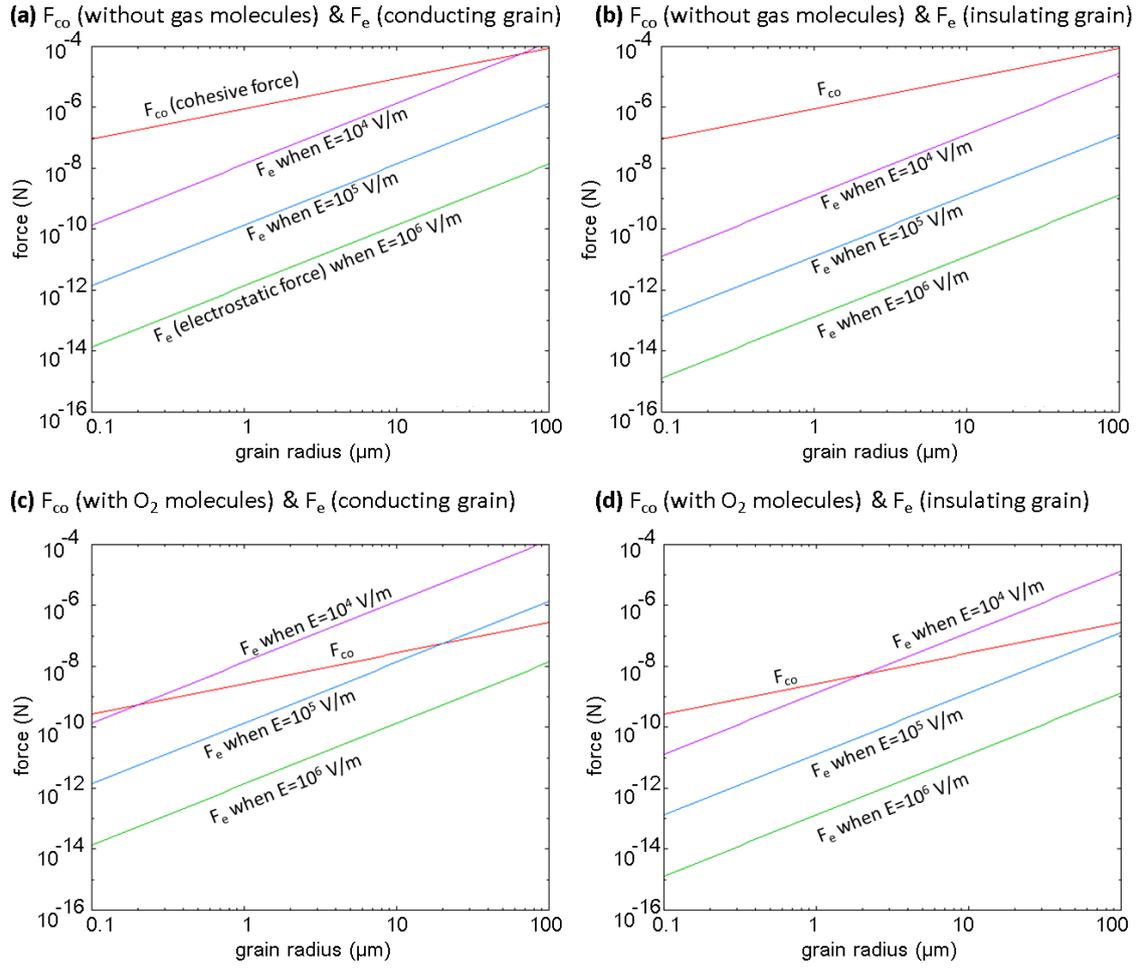

**Figure 3.** Cohesive ($F_{co}$) and electrostatic forces ($F_e$) acting on dust grains as a function of grain size. (a) $F_{co}$ and $F_e$ acting on conducting grains without gas molecules. (b) $F_{co}$ and $F_e$ acting on insulating grains without gas molecules. (c) $F_{co}$ and $F_e$ acting on conducting grains with adsorbed $O_2$ molecules. (d) $F_{co}$ and $F_e$ acting on insulating grains with adsorbed $O_2$ molecules. We assumed $\gamma = 5.8 \times 10^{-4}$ J·m$^{-2}$ and 0.19 J·m$^{-2}$ for ring particles with and without $O_2$ molecule, respectively. Here we show $F_e$ in three cases: $E = 3.0 \times 10^4$ V/m, $3.0 \times 10^5$ V/m, and $3.0 \times 10^6$ V/m. Note that $F_{co}$ and $F_e$ are described in Eq. (1), (5), and (11).

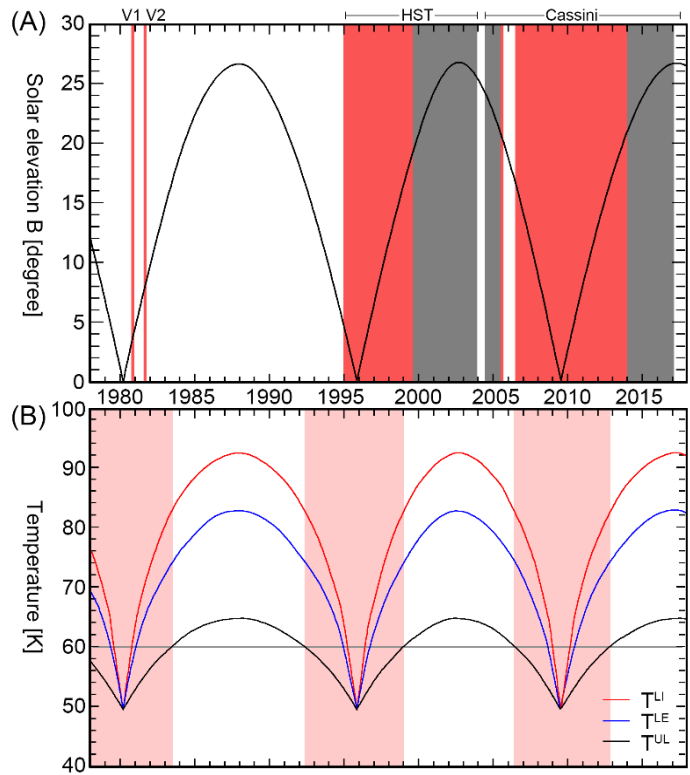

**Figure 4.** Comparison of the observed appearance of the spokes in Saturn's rings (A) and the prediction based on our model (B). (A) The black line represents the absolute value of the ring opening angles as a function of time. Red lines and red areas mean the season when spokes were visible. The gray areas show the periods when spokes were invisible by HST or Cassini. This figure has been created by the modification of Fig. 16.8 of Horányi et al. (2009), which shows the seasons when the spokes appear as a function of time, combining Voyager flybys in 1980 and 1981, Hubble Space Telescope observations during 1994-2004, and early Cassini observations by 2009. In addition, by examining Cassini images from 2013 to 2016, we identify spokes in the images obtained in Jan. 5, 2014 (W1767601332-5647), while no spokes have been identified since then. Based on these observations, we create this figure. (B) Seasonal temperature variations of the B ring based on the thermal modeling (Spilker et al. 2013, Ferrari and Reffet 2013) and the prediction of spoke appearance based on our model together with this temperature. Red, blue, and black lines represent the temperature of the lit face of the ring at ingress ($T^{LI}$) and egress ($T^{LE}$) of the planetary shadow, and that of the unlit face ($T^{UL}$), respectively. Red-shaded areas indicate the seasons of spoke appearance predicted from our model, i.e., when the lowest temperature of the ring becomes below 60 K.

**Appendix A. Cohesive force acting on dust grains**

        Forward-scattering observations during radio or stellar ring occultations reported that the main ring consists of particles ranging from 1 cm to 10 m in radius and lacked mm-sized or μm-sized dust particles floating in free space (Colwell et al. 2010, French and Nicholson 2000, Marouf et al. 1983, Zebker et al. 1985). For example, a ground-based observation suggests that the minimum radius of ring particles is 1 cm for the A ring, 0.1 cm for the Cassini Division, 30 cm for the B ring, and 1 cm for the C ring (French and Nicholson 2000). Note that these sizes represent the size of individual ring particles if the particles are freely floating, while they should be regarded as the size of aggregates if ring particles form aggregates. Spectral reflectance studies of ring particles in ultraviolet, visible, and infrared wavelength ranges reported that ring particles are covered with grains with a radius of 5–20 μm (Nicholson et al. 2008, Poulet et al. 2003, Doyle et al. 1989, Bradley et al. 2010). Thermal inertia of the main rings is consistent with ring particles with a dusty regolith layer (Morishima et al. 2011). Hence, these seem to indicate that the surface of meter to cm-sized ring particles are covered with μm-sized dust grains, or that meter to cm-sized ring particles are aggregates of μm-sized dust grains.

        Based on the so-called JKR theory, Bodrova et al. (2012) developed a model where μm-sized dust grains stick to cm-sized ring particles owing to cohesive forces. The requirement that a dust grain is able to stick adhesively to other ring particles depends on the relative velocity at collision, the size of the dust grain, and its surface energy. The maximum relative velocity for adhesive sticking between a spherical grain and a flat surface is given by (Chockshi et al. 1993, Dominik and Tielens 1997, Kimura et al. 2015)

$$v_{stick} = \left(\frac{27 c_1 \pi^{2/3}}{2^{5/3}}\right)^{1/2} \left(\frac{\gamma^5 (1-\nu^2)^2}{a^5 \rho^3 E_Y^2}\right)^{1/6}, \quad (A1)$$

where $c_1$ is a constant ($c_1 = 1$) (Thornton and Ning 1998, Brilliantov et al. 2007, Chockshi et al. 1993), $\gamma$ is the surface energy, $\nu$ is Poisson's ratio, $a$ is the radius of the grain, $\rho$ is the density of the grain, and $E_Y$ is Young's modulus. Transforming this formula, we can estimate the minimum radius of particles that is able to escape from the surface of other larger particles (i.e., the maximum radius of particles that is able to stick the surface of other larger particles owing to cohesion) to be

$$a = 5.71 \left(\frac{\gamma^5 (1-\nu^2)^2}{v_{stick}^6 \rho^3 E_Y^2}\right)^{1/5}. \quad (A2)$$

In general, smaller particles stick more easily (Figure A1). As an example, here we assume an icy particle with $\gamma = 0.190$ J·m$^{-2}$ (Gundlach et al. 2011), $\nu = 0.32$ and $E_Y = 9.3$ GPa (Gammon et al. 1983), and $\rho = 900$ kg·m$^{-3}$. For simplicity, here we estimate the relative velocity between ring particles from their velocity dispersion (Bodrova et al. 2012, Salo 1995, Ohtsuki and Emori 2000); a more detailed discussion based on N-body simulation can be found in Ohtsuki et al. (2020). In dilute rings such as the C ring, the velocity dispersion is determined by mutual collisions between particles and is proportional to the particle radius. In dense rings such as the A and B rings, on the other hand, small-scale structures called gravitational wakes are formed and the velocity dispersion in this case is proportional to the surface density of the ring, although the relative velocity between particles in the same wake can be smaller than the velocity dispersion (Salo 1995). From the results of dynamical studies (Bodrova et al. 2012, Salo 1995, Ohtsuki and Emori 2000) and the surface densities estimated from observations and numerical simulations (Porco et al. 2008, Colwell et al. 2009, Robbins et al. 2010), we estimate the typical impact velocity to be 1 ~ 2 mm/s for the A ring, 6 ~ 12 mm/s for the B ring, and 0.5 ~ 1 mm/s for the C ring, respectively. From these numbers, we obtain $a$= 3.1~ 7.2 mm for the A ring, $a$ = 0.36~0.84 mm for the B ring, and $a$ = 7.2 ~ 17 mm for the C ring (Figure A1). This is consistent with the deficiency of mm-sized or μm-sized particles freely floating in the main ring.

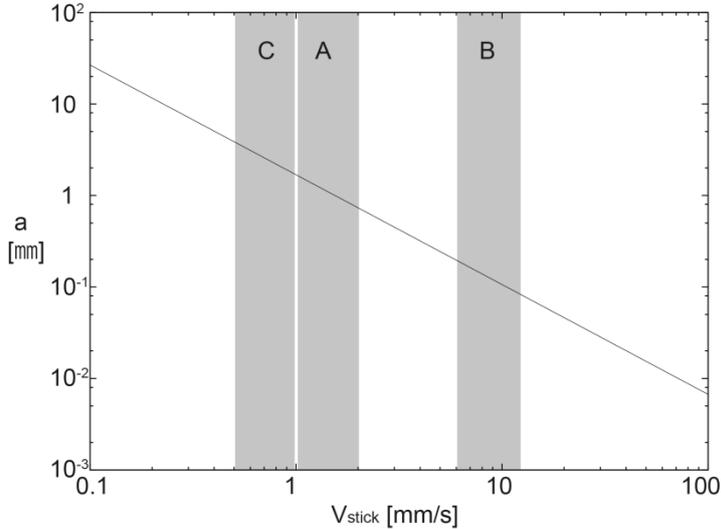

**Figure A1.** The maximum radius of particles ($a$) which is able to stick to the surface of other larger particles owing to cohesion, as a function of the relative velocity between particles ($v_{stick}$). Grey regions indicate typical relative velocities of ring particles estimated from their velocity dispersions in the A, B, and C rings, respectively.

### Appendix B. Conductivity of ring particles

The conductivity of dust grains is one of the essential properties of grains when considering their detachment (Sow et al. 2013, von Holstein-Rathlou et al. 2012), although previous models on spoke formation seem to have overlooked this issue. At the same time, the electric conductivity of icy particles is complicated because it significantly depends on the temperature, minor compartments, and the structural and dynamical properties of the surface of the icy particles, although water ice is an insulating material.

Experimental studies on electrostatic lofting of submillimeter to millimeter-sized particles have revealed that insulating grains covered with electrically conductive molecules behave like conductors (Sow et al. 2013, von Holstein-Rathlou et al. 2012). Therefore, we may expect that water-ice grains covered with electrically conductive oxides can be detached from their sticking surfaces, owing to the enhancement of electrical currents on the oxygen-doped surfaces of water ice grains by proton jumping. The interaction between $O^+$ ions and $H_2O$ molecules in air is known to be a reaction for proton transfer, since the ionization energy of the former (13.62 eV) is higher than that of the latter (12.62 eV) (Hansel et al. 1995). Although $O_2^+$ ions do not interact with $H_2O$ molecules in air, owing to the lower ionization energy of $O_2^+$ ions (12.06 eV), a proton transfer reaction most likely takes place between $O_2^+$ ions and $H_2O$ ice grains, because of the lower work function (8.7 eV) of $H_2O$ ice (Baron et al. 1978). In Saturn's rings, it is, therefore, natural to assume that the adsorption of $O^+$ and $O_2^+$ ions on the surface of water-ice grains converts electric properties of the grains into conductors. In any case, the conductivity of ring particles has not been measured yet and this issue is largely an open question at present.

**Appendix C. Recent observations of the lunar horizon glow.**

We have received a comment that the so-called lunar horizon glow (LHG) was not confirmed by modern spacecraft observations, such as the Lunar Reconnaissance Orbiter (Barker et al. 2019) and Chang'E-3 (Li et al. 2020). Here we stated our view as follows.

The LHG has been first discovered by the Surveyor 7 spacecraft (Gault et al, 1968), and the Lunar Ejecta and Meteorite experiment set on the lunar surface during the Apollo 17 mission have detected the falls of these hovering grains (Berg et al. 1976). As on the Moon, this process has been proposed on asteroids or satellites (e.g. Lee 1996, Colwell et al. 2005, Senshu et al. 2015, Hirata and Miyamoto 2012). Most reliable observations for LHG is the LEAM (Apollo 17 surface instrument) observations. The instrument recorded high flux of dust particle impacts during sunrise and sunset terminator crossings (Berg et al. 1976). In addition, the instrument detected highly charged, slowly moving grains. The fact that grains are highly charged is consistent with the view that dust grains are electrostatically lofted, because the grains gain high charge in the progress prior to its electrical levitation. Also, the direction of these grains detected indicates the grains moving from dayside to nightside. These properties are makes no sense under ejecta or interplanetary impacts. The LEAM detected the mass flux of floating dust grains with $4\times10^{-18}$ g cm$^{-2}$ s$^{-1}$(Berg et al. 1976). It corresponds to $3.4\times10^{-13}$ g cm$^{-2}$ even during 24 hours.

On the other hand, the lower limit of detectable column mass of dust is $10^{-11}$ g cm$^{-2}$ in the case of instrument onboard LRO (Barker et al. 2019). Note that, because this lower limit assumes 0.1 μm as a typical dust grain size and it is known that LHG dust has the order of 1 μm in size, the lower limits of LRO's observation must be much larger than $10^{-11}$ g cm$^{-2}$. Also, the lower limit of detectable column mass of dust is $2.46\times10^{-6}$ g cm$^{-2}$ in the case of the Chang'E-3 mission (Li et al. 2020). Therefore, instrument abilities of the LRO and Chang'E-3 are insufficient to detect LHG observed by Apollo 17, and therefore there is no contradiction between the observations of the Apollo missions and the results of LRO and Chang'E-3. In addition, we would like to note that, one of modern spacecraft, the Lunar Dust Experiment (LDEX) onboard the LADEE spacecraft (The Lunar Atmosphere and Dust Environment Explorer), reported five dust enhancement events which happen near a twilight crater with dust densities comparable to the Apollo measurements (Xie et al. 2020).

## Appendix D. The charge patch model and effect of cohesion

The so-called charge patch model (CPM) has been first proposed by Wang et al. (2016), where dust release takes place without intense electric fields across the terminator demonstrated in Section 3. The mechanism based on the CPM is as follows; (i) a dusty surface is not flat plane but consisting of microscopic irregular structures formed by the shape of each grains, and therefore, a space like a cave, so-called micro-cavity, frequently appears beneath a dust grain (Figure D1), (ii) when UV or plasma irradiates the dusty surface, it hardly hits micro-cavities but photoelectrons or secondary electrons emitted by it easily hits micro-cavities, (iii) these micro-cavities tend to be charged up negatively, and (iv) then, the Coulomb repulsion within the cavity lifts a dust grain in the top. The CPM has been further advanced by Schwan et al. (2017) and Orger et al. (2018) theoretically and Orger et al. (2021) experimentally. Also, they conclude that dust grains are levitated by UV or plasma irradiation alone.

However, all of their experiments, Wang et al. (2016) and Orger et al. (2021), do not seem to have exactly simulated the environment of the LHG nor provided evidence that is consistent with observational results of the LHG. Their experiments have only ever been performed in moderate vacuum conditions of $10^{-4}$ Pa or $4 \times 10^{-3}$ Pa and with the use of hydrophilic silicate particles that are not ablated. Therefore, it is likely that the surface of the particles is covered by water molecules. The existence of water molecules on the surface is known to reduce the cohesion of hydrophilic silicate particles significantly (e.g., Kimura et al. 2014, 2015, Steinpilz et al. 2019). The cohesion between dust grains in their experiments was negligibly small, and therefore their experiments have been likely conducted under conditions where dust emission was much easier to occur than on actual lunar surface outside the terminator region, in other words, dust grains utilized in Wang et al. are much looser than the actual lunar soil. In fact, Wang et al. (2016) noted that the cohesive force acting on 10-μm dust grains utilized in their experiment amounts to $10^{-8}$ N, although the paper did not explicitly note the surface energy of dust grains utilized. The cohesion with $10^{-8}$ N indicates the grain having very small surface energy, $\gamma=10^{-4}$ J m$^{-2}$, which is obtained from Eq. (1). This is three-to-four orders of magnitude lower than the surface energy of silicates in vacuum, $\gamma= 0.275$ J m$^{-2}$ for amorphous silica (experimentally determined by Tarasevich, 2007) and $\gamma= 1.5$ J m$^{-2}$ for crystalline silica. This seems to indicate that the dust grains in their experiments are covered by gas molecules. Since their experiment was carried out in a low-vacuum chamber with $10^{-4}$Pa, a large number of gas molecules, which play a vital role in reducing surface energy of dust grains, would have inevitably remained adsorbed onto the surface of dust grains. In addition, it seems that it was difficult for them to accomplish complete ablation of gas molecules on the surface of dust grains by evacuation, because Mars regolith simulants utilized in their experiment contain a considerable amount of water (Allen et al. 1998). If their model is actually at work and accounts for their experiments correctly, then the electrostatic lofting of dust grains should be common phenomena, but in reality it is a rare phenomenon that has never been observed outside the terminator region.

Furthermore, in their experiments, a direct link between the presence of micro-cavities and dust release observed is yet to be demonstrated. One could simply account for the experimental results with the reduction of cohesion by water molecules and fluffy agglomeration of particles as demonstrated by Kimura et al. (2015). It is known that one of the essential issues to generate the dust release from the surface of airless bodies is how to manage the strong cohesion. However, their experiment has been performed on Earth likely under the influence of reduced cohesion, as we discussed above. Therefore, we think that UV radiation and/or plasma charging alone are not able to overcome the cohesion acting on actual dust grains on the Moon, unless accompanied by a significant reduction in the surface energy of grains. It should be noted that our model does not contradict with their experiments at all, because the experimental results can be explained by the effect of cohesion as we described above.

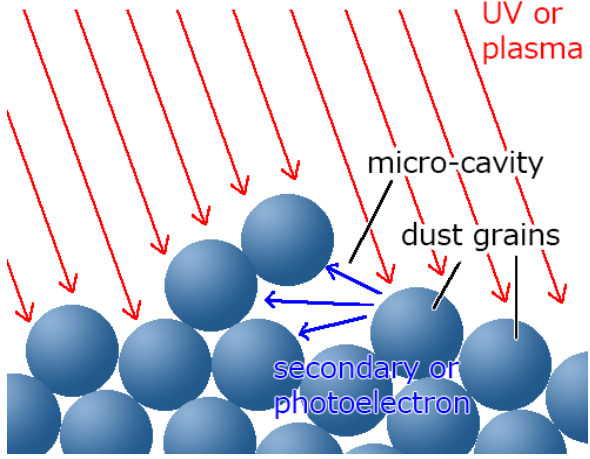

**Figure D1.** On a dusty surface, a space like a cave frequently appears beneath dust grains. Such micro-cavities are not exposed by UV or plasma but exposed by photoelectrons or secondary electrons (blue arrows) emitted by UV irradiation or plasma absorption (red arrows). The walls of the micro-cavity will be charged up negatively, although some of the walls would be charged up positively. The Coulomb repulsive forces between the roof and the floor of the cavities act on the dust grain above the cavity.

**Appendix E. An amount of O₂ molecules on ring particles.**

Although it is difficult to estimate an amount of $O_2$ molecules on ring particles, we attempt to estimate it in this section. According to Tseng et al. (2010), the mean free time of an oxygen molecule to collide with a ring particle is $10^4$ seconds (=2.7 hours). The passage of Saturn's shadow takes 2 hours. Below 60 K in ice temperature, over 90% of oxygen molecules are adsorbed onto ice in collision (He et al. 2016). Therefore, majority of $O_2$ molecules are adsorbed onto ring particles at the egress of Saturn's shadow.

Although it is difficult to estimate the surface area of a particle in Saturn's ring, we made the following estimates. French and Nicholson (2000) estimates the size distribution of ring particles per unit area of ring plane (particles m⁻² m⁻¹) in the B ring to be

$n = ar^{-2.75}$ (0.30 m ≤ $r$ ≤ 20 m) , (E1)

where $r$ is the radius of particle and $a$ is the number of $n$ when $r = 1$. Note that the B ring lacks particles smaller than 0.30 m and larger than 20 m (French and Nicholson 2000). The total surface area of ring particles in the B ring per unit area (1 m²) of ring plane (m² m⁻²) is given by

$S = \int_{0.3}^{20} 4\pi r^2 \cdot ar^{-2.75} \, dr = 69a$ , (E2)

if we assume ring particles are sphere. The value of $a$ is difficult to estimate because of optically thickness of the B ring. However, it is known that the B-ring's typical surface mass density is roughly 1000 kg m$^{-2}$ (Hedman and Nicholson 2016) and the bulk density of ring particles are 500 kg m$^{-3}$, and therefore, the total volume of B-ring's particles per unit area of ring plane is $V = 2.0$ m$^3$ m$^{-2}$. The total volume of ring particles per unit area of ring plane (m$^3$ m$^{-2}$) can be also estimated from French and Nicholson's size distribution:

$$V = \int_{0.3}^{20} \frac{4\pi}{3} r^3 \cdot ar^{-2.75} \, dr = 141a \quad . \text{(E3)}$$

Therefore, we can obtain $a = 0.014$ and therefore $S = 0.97$ m$^2$ m$^{-2}$. As a result, we can obtain $S = 1$ cm$^2$ cm$^{-2}$ as a total of the surface area of ring particles per 1 cm$^2$ of ring plane. This is roughly consistent with the fact that the optical depth of the B ring is on the order of unity.

The ring atmosphere is generated by photo-sputtering of icy ring particles (Johnson et al. 2006) and, as a consequence, its amount is at minimum near the equinox and at maximum near the solstice (Tseng et al. 2010). According to Tseng et al. (2010), the column density of O$_2$ molecules per 1 cm$^2$ of ring plane is on the order of 10$^{11}$ molecular cm$^{-2}$ near the equinox and on the order of 10$^{12}$ molecular cm$^{-2}$ near the solstice. There is a seasonal variation of about one order of magnitude. If we assume the column number density of O$_2$ molecules to be 10$^{12}$ cm$^{-2}$ and these molecules are distributed homogenously on the surface of ring particles, the surface number density of O$_2$ molecules on the surface of ring particles would be 10$^{-2}$ nm$^{-2}$. The average nearest neighbor distance of O$_2$ molecules is 10 nm, and therefore, basically this is monolayer. We mention again here that dust release would be rather easily when the amount of adsorbed gas molecules is small; for example, it is known that the very low surface energy of dry sands in dry air results from the small size of the capillary neck formed by a low amount of gas molecules (Vigil et al. 1994). On the other hand, this estimate has a great uncertainty because there are areas that are easily hit by oxygen molecules and areas that are not, ring particles actually form aggregates, and there should be accumulated amounts of O$_2$ molecules on or in particles from the past.

## Appendix F. Why the spokes do not appear in optical thin zones

According to Spilker et al. (2013), not only the B ring but also the Cassini Division and A ring become lower than 60 K in temperature near the equinox. For example, the temperature of the A ring becomes lower than 60K for the ring opening angle, $B < 3$ degree (Morishima et al. 2016). Therefore, we expect that the dust release from ring particles itself should occur in the A ring then. However, we cannot identify any reliable examples of the spokes in the A ring in images taken by the Voyager and Cassini.

It has been known that the radial location of the observed spokes is correlated with the optical depth of the background rings (Grün et al. 1992). In detail, the spokes predominantly appear in the central B ring (99000km to 110000 km in radial distance from Saturn's center), which is the optically thickest zone in the B ring (red arrow in Figure F1). However, the B ring is relatively optically thin in its inner and outer zones, and the spokes barely appear there. Even in the central thick zone of the B ring, the ring plane has many intermittent optically thin zones, and the spokes do not appear there (blue arrow in Figure F1). In short, the radial location of the spokes is strongly correlated with the local optical depth of the rings. In fact, the optical depth of the A ring is close to that of the inner B ring. We may, therefore, attribute the non-detection of spokes in the A ring to its small optical depth compared with the central B ring.

The optical depth is related to (i) the production rate of $O_2$, (ii) the amount of dust grains, and/or (iii) the visibility of the spokes itself. (i) First, the amount of molecular $O_2$ and the optical depth of the background rings show a positive correlation, because the $O_2$ is generated by photo-dissociation or sputtering of icy ring particles (Tokar et al. 2005). Therefore, the amount of $O_2$ molecule in the A ring is most likely smaller than that in the B ring. Unfortunately, the amount of $O_2$ in the A ring is qualitatively uncertain (Young et al. 2005), and therefore we cannot argue this idea conclusively. (ii) Second, the optical depth would be correlated with a potential amount of dust grains. Even if the dust release occurs, the amount of freely floating dust grains may be insufficient to exceed the detection limit of the Cassini spacecraft or the Hubble space telescope. (iii) Third, at a low phase angle, the optically thinner ring is darker (the spokes originally look dark), while at a high phase angle, the optically thinner ring is brighter (the spokes originally look bright). Therefore, the contrast between the spokes and the background optically thinner rings is much smaller (Grün et al. 1992). As a result, the spokes would be hardly visible in optically thin rings.

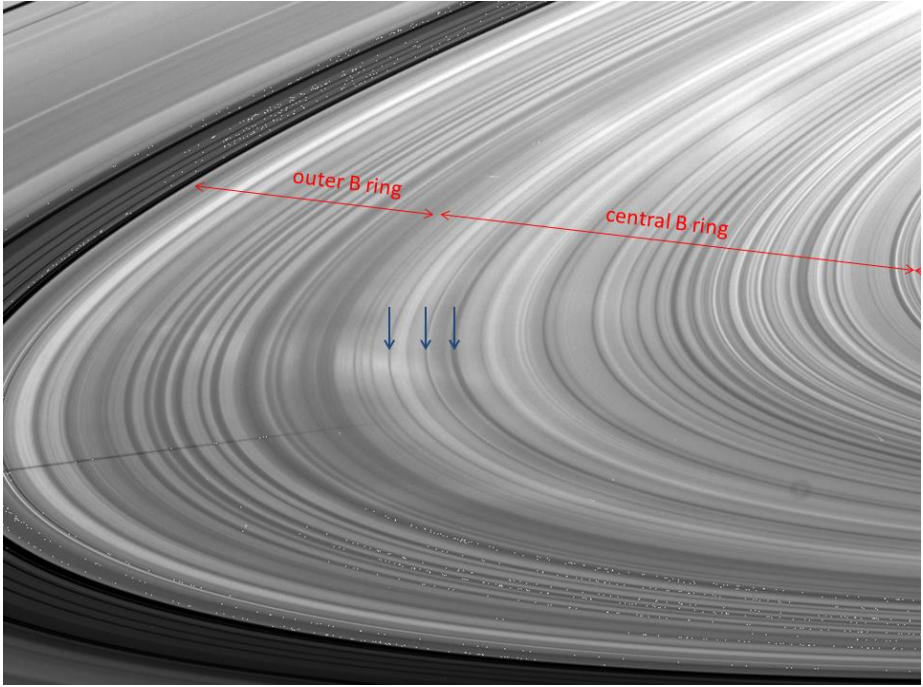

**Figure F1.** The spokes and the B ring (N1631413635).

**Appendix G. Possible explanation for an intermittent appearance of spokes.**

An intermittent appearance of spokes may be attributed to the presence of already freely-floating charged dust grains in the ring, rather than the detachment process of dust grains adsorbed on ring particles. The absence of the spokes is not necessarily ascribed to no detachment of dust grains from the surface of ring particles, because the detection of a spoke by space-borne optical instruments depends on the altitude of dust grains from the ring plane and the spatial density of the grains. We briefly examine the dynamics of spoke-forming dust grains after the detachment of the grains, although detailed numerical simulation on the grain dynamics is beyond the scope of this paper. Since the level of spoke activity (i.e., the integral of optical depth for each spoke) has a period equal to that of the Saturn's Kilometric Radiation (Porco and Danielson 1982, Mitchell et al. 2013), the altitude and density of a spoke might be associated with a temporal variation in field-aligned electric currents, even though our model does not explain the relation between the spokes and the Saturn's Kilometric Radiation. As proposed by Ip and Mendis (1983), electric charging of ring particles would generate field-aligned Birkland currents flowing into the ring at dawn and from the ring at dusk. If field-aligned Birkland currents lift up negatively freely-floating dust grains above the ring plane enough to be observed, then the upward motion of negatively charged grains at dawn may largely disturb or even halt the already existing Birkland currents at dawn. The disturbance in the Birkland currents suppresses the visibility of the spokes, and may explain their observed intermittent appearance.

## Appendix H. The ultraviolet flux from Saturnshine

Ring particles at the equinox are barely illuminated by solar UV radiation. Near equinox, the UV photons reflected by Saturn (i.e., Saturnshine) would play a major role in the charging mechanism of the particles. The total solar UV flux for the full disk of Saturn is given by $\pi R^2 F_{sun}$, where $R$ is the Saturn's radius and $F_{sun}$ is the solar UV flux per unit area at Saturn's orbit. The total UV flux reflected from the entire Saturn is given by $a\pi R^2 F_{sun}$, where $a$ is the albedo of Saturn in the ultraviolet.

If the Saturnshine is uniformly emitted from the entire surface of Saturn ($=4\pi R^2$), the total UV flux reflected from Saturn per unit surface area of Saturn ($F_{reflect}$) is obtained by dividing $a\pi R^2 F_{sun}$ by $4\pi R^2$. Then, we obtain

$$F_{reflect} = \alpha F_{sun}/4 \ . \text{(H1)}$$

In nature, the Saturnshine is emitted intensively from the dayside of Saturn and barely from the nightside. For the sake of simplicity, we assume uniform Saturnshine here. Eq. (H1) is regarded as an average value of UV flux through one orbital cycle of ring particles. At noon ansa of the ring plane, the Saturnshine UV flux would be a couple of times greater than Eq. (H1). Based on simple geometrical consideration, the Saturnshine UV flux received by a ring particle at a distance ($r$) from Saturn's center is given by

$$F_{saturnshine\ at\ ring} = sI\Omega \ , \text{(H2)}$$

where $I$ is the intensity of Saturnshine and $\Omega$ is the solid angle subtended by Saturn, which are described by

$$I = F_{reflect}/\pi \text{ (H3)}$$

and

$$\Omega = 2\pi \left(1 - \sqrt{1 - (R/r)^2}\right). \text{(H4)}$$

The value of $s$ in Eq. (H2) is the shadowing factor of Saturnshine due to other ring particles ($0 < s < 1$); $s = 1$ for a ring particle not shadowed by other ring particles (i.e., the ring particle receives Saturnshine without any screening), while $s = 0$ for a ring particle completely shadowed by other particles. An actual ring particle is more or less shadowed by many other particles. Particles in the uppermost layer of the ring plane at the noon ansa is able to see the upper half of Saturn, and then $s = 0.5$. As a result, we obtain

$$F_{saturnshine\ at\ ring} = \frac{1}{2}\alpha s \left(1 - \sqrt{1-(R/r)^2}\right) F_{sun} \quad . \text{(H5)}$$

Because the work function of ice is 8.7 eV (Baron 1978), and therefore, far ultraviolet radiation with a wavelength < 282 nm contributes to photoelectron emission from icy ring particles. Saturn could reflect roughly 50% of ultraviolet radiation in the wavelength of 180-282 nm (Clarke 1982). Therefore, assuming $\alpha$=0.5, $s$=0.5, $r$ = 1.0x10$^5$km, and $R$=5.8x10$^4$km, we obtain $F_{saturnshine\ at\ ring}$ = 0.023$F_{sun}$. This indicates that Saturnshine UV flux is comparable with the solar UV flux ($F_{sun}\sin B$) when $B$= 1.3 degree. In other words, during the season of $B$< 1.3 degree, Saturnshine dominates the UV flux onto ring particles.

# References

Allen, C. C. et al. (1998), MARTIAN REGOLITH SIMULANT JSC MARS-1, abstract #1690 in Lunar and Planetary Science Conference XXIXLPSC abstract, Houston, TX.

Arp, P. A. and S. G. Mason (1977) Particle behaviour in shear and electric fields. *Colloid and Polymer Science,* 255, 566-584.

Barker, M.K., E. Mazarico, T.P. McClanahan, X. Sun, G.A. Neumann, D.E. Smith, M.T. Zuber, J.W. Head (2019) Searching for Lunar Horizon Glow With the Lunar Orbiter Laser Altimeter. *Journal of Geophysical Research Planets* 124, Issue11, 2728-2744.

Baron, B., D. Hoover and F. Williams (1978) Vacuum ultraviolet photoelectric emission from amorphous ice. *The Journal of Chemical Physics,* 68, 1997-1999.

Berg, O., H. Wolf, and J. Rhee (1976). 2.2.4 Lunar Soil Movement Registered by the Apollo 17 Cosmic Dust Experiment. International Astronomical Union Colloquium, 31, 233-237. doi:10.1017/S0252921100051757

Bodrova, A., J. Schmidt, F. Spahn and N. Brilliantov (2012) Adhesion and collisional release of particles in dense planetary rings. *Icarus,* 218, 60-68.

Bradley, E. T., J. E. Colwell, L. W. Esposito, J. N. Cuzzi, H. Tollerud and L. Chambers (2010) Far ultraviolet spectral properties of Saturn's rings from Cassini UVIS. *Icarus,* 206, 458-466.

Brilliantov, N. V., N. Albers, F. Spahn and T. Pöschel (2007) Collision dynamics of granular particles with adhesion. *Physical Review E,* 76, 051302.

Cadenhead, D. A., M. G. Brown, D. K. Rice and J. R. Stetter (1977) Some surface area

and porosity characterizations of lunar soils. *Proceedings of 8th Lunar and Planetary Science Conference*, 1291-1303.

Chockshi, A., A. G. G. Tielens and D. Hollenbach (1993) Dust coagulation. *The Astrophysical Journal,* 407, 806-819.

Ciarniello, M., G. Filacchione, E. D'Aversa, F. Capaccioni, P. D. Nicholson, J. N. Cuzzi, R. N. Clark, M. M. Hedman, C. M. Dalle Ore, P. Cerroni, C. Plainaki and L. J. Spilker (2019) Cassini-VIMS observations of Saturn's main rings: II. A spectrophotometric study by means of Monte Carlo ray-tracing and Hapke's theory. *Icarus,* 317, 242-265.

Clark, R. N. (2009) Detection of Adsorbed Water and Hydroxyl on the Moon. *Science,* 326, 562-564.

Clarke, J., H. Moos and P. Feldman (1982) The far-ultraviolet spectra and geometric albedos of Jupiter and Saturn. *The Astrophysical Journal,* 255, 806-818.

Colwell, J. E., A. A. S. Gulbis, M. Horányi and S. Robertson (2005) Dust transport in photoelectron layers and the formation of dust ponds on Eros. *Icarus,* 175, 159-169.

Colwell, J. E., L. W. Esposito, R. G. Jerousek, M. Sremčević, D. Pettis and E. T. Bradley (2010) Cassini UVIS Stellar Occultation Observations of Saturn's Rings. *The Astronomical Journal,* 140, 1569.

Colwell, J. E., P. D. Nicholson, M. S. Tiscareno, C. D. Murray, R. G. French and E. A. Marouf (2009) The Structure of Saturn's Rings. In *Saturn from Cassini-Huygens,* eds. M. K. Dougherty, L. W. Esposito and S. M. Krimigis, 375-412. Dordrecht: Springer Netherlands.

Colwell, J. E., S. Batiste, M. Horányi, S. Robertson and S. Sture (2007) Lunar surface: Dust dynamics and regolith mechanics. *Reviews of Geophysics,* 45, RG2006.

Criswell, D. R. (1973) Horizon-Glow and the Motion of Lunar Dust. In *Photon and Particle Interactions with Surfaces in Space: Proceedings of the 6th Eslab Symposium, Held at Noordwijk, the Netherlands, 26–29 September, 1972,* ed. R. J. L. Grard, 545-556. Dordrecht: Springer Netherlands.

Criswell, D. R. and B. R. De (1977) Intense localized photoelectric charging in the lunar sunset terminator region, 2. Supercharging at the progression of sunset. *Journal of Geophysical Research,* 82, 1005-1007.

D'Aversa, E., G. Bellucci, P. D. Nicholson, M. M. Hedman, R. H. Brown, M. R. Showalter, F. Altieri, F. G. Carrozzo, G. Filacchione and F. Tosi (2010) The spectrum of a Saturn ring spoke from Cassini/VIMS. *Geophysical Research Letters,* 37, L01203.

De, B. R. and D. R. Criswell (1977) Intense localized photoelectric charging in the lunar sunset terminator region, 1. Development of potentials and fields. *Journal of Geophysical Research,* 82, 999-1004.

Dominik, C. and A. G. G. M. Tielens (1997) The Physics of Dust Coagulation and the Structure of Dust Aggregates in Space. *The Astrophysical Journal,* 480, 647.

Doyle, L. R. and E. Grün (1990) Radiative transfer modeling constraints on the size of the spoke particles in Saturn's rings. *Icarus,* 85, 168-190.

Doyle, L. R., L. Dones and J. N. Cuzzi (1989) Radiative transfer modeling of Saturn's Outer B ring. *Icarus,* 80, 104-135.

Eplee, R. E. and B. A. Smith (1984) Spokes in Saturn's rings: Dynamical and reflectance properties. *Icarus,* 59, 188-198.

Farrell, W. M., M. D. Desch, M. L. Kaiser, W. S. Kurth and D. A. Gurnett (2006) Changing electrical nature of Saturn's rings: Implications for spoke formation. *Geophysical Research Letters,* 33, L07203.

Farrell, W. M., W. S. Kurth, D. A. Gurnett, A. M. Persoon and R. J. MacDowall (2017) Saturn's rings and associated ring plasma cavity: Evidence for slow ring erosion. *Icarus,* 292, 48-53.

Ferrari, C. and E. Reffet, E. (2013) The dark side of Saturn's B ring: Seasons as clues to its structure. Icarus 223, 28-39, doi:http://dx.doi.org/10.1016/j.icarus.2012.12.006 (2013).

Flanagan, T. M. and J. Goree (2006) Dust release from surfaces exposed to plasma. *Physics of Plasmas,* 13, 123504.

French, R. G. and P. D. Nicholson (2000) Saturn's Rings II: Particle Sizes Inferred from Stellar Occultation Data. *Icarus,* 145, 502-523.

Gammon, P., H. Kiefte and M. Clouter (1983) Elastic constants of ice samples by Brillouin spectroscopy. *The Journal of Physical Chemistry,* 87, 4025-4029.

Gault, D. et al. (1968) Post-sunset horizon glow. In Surveyor Project Final Report Part 2, Science Results, JPL technical report 32-1265, 401-405.

Goertz, C. K. and G. Morfill (1983) A model for the formation of spokes in Saturn's ring. *Icarus,* 53, 219-229.

Grün, E., C. K. Goertz, G. E. Morfill and O. Havnes (1992) Statistics of Saturn's spokes. *Icarus,* 99, 191-201.

Grün, E., G. E. Morfill, R. J. Terrile, T. V. Johnson and G. Schwehm (1983) The evolution of spokes in Saturn's B ring. *Icarus,* 54, 227-252.

Gundlach, B., S. Kilias, E. Beitz and J. Blum (2011) Micrometer-sized ice particles for planetary-science experiments – I. Preparation, critical rolling friction force, and specific surface energy. *Icarus,* 214, 717-723.

Hansel, A., A. Jordan, R. Holzinger, P. Prazeller, W. Vogel and W. Lindinger (1995) Proton transfer reaction mass spectrometry: on-line trace gas analysis at the ppb level. *International Journal of Mass Spectrometry and Ion Processes,* 149-150, 609-619.

Hartzell, C. M. and D. J. Scheeres (2011) The role of cohesive forces in particle launching on the Moon and asteroids. *Planetary and Space Science,* 59, 1758-1768.

He, J., K. Acharyya and G. Vidali (2016) Sticking of Molecules on Nonporous Amorphous Water Ice. *The Astrophysical Journal,* 823, 56.

Hedman, M.M., and P.D.Nicholson (2016) The B-ring's surface mass density from hidden density waves: Less than meets the eye? *Icarus* 279, 109-124.

Hibbitts, C. A., G. A. Grieves, M. J. Poston, M. D. Dyar, A. B. Alexandrov, M. A. Johnson and T. M. Orlando (2011) Thermal stability of water and hydroxyl on the surface of the Moon from temperature-programmed desorption measurements of lunar analog materials. *Icarus,* 213, 64-72.

Hill, J. R. and D. A. Mendis (1981) On the braids and spokes in Saturn's ring system. *The moon and the planets,* 24, 431-436.

Hirata, N., and H. Miyamoto (2012) Dust levitation as a major resurfacing process on the surface of a saturnian icy satellite, Atlas. *Icarus* 220, 106-113.

Horanyi, M., G. E. Morfill and T. E. Cravens (2010) Spokes in Saturn's B Ring: Could Lightning be the Cause? *IEEE Transactions on Plasma Science,* 38, 874-879.

Horányi, M., J. A. Burns, M. M. Hedman, G. H. Jones and S. Kempf (2009) Diffuse Rings. In *Saturn from Cassini-Huygens,* eds. M. K. Dougherty, L. W. Esposito and S. M. Krimigis, 511-536. Dordrecht: Springer Netherlands.

Hsu, H.-W. et al. (2018) In situ collection of dust grains falling from Saturn's rings into its atmosphere. *Science,* 362, eaat3185.

Ip, W. H. (1995) The Exospheric Systems of Saturn's Rings. *Icarus,* 115, 295-303.

Ip, W. H. and D. A. Mendis (1983) On the equatorial transport of Saturn's ionosphere as driven by a dust-ring current system. *Geophysical Research Letters,* 10, 207-209.

Israelachvili, J. N. (2011) Adhesion and Wetting Phenomena. In *Intermolecular and Surface Forces (Third Edition)*, 415-467, San Diego: Academic Press.

Johari, G. P. and E. Whalley (1981) The dielectric properties of ice Ih in the range 272–133 K. *The Journal of Chemical Physics,* 75, 1333-1340.

Johnson, K. L., K. Kendall and A. D. Roberts (1971) Surface Energy and the Contact of Elastic Solids. *Proceedings of the Royal Society of London. A. Mathematical and Physical Sciences,* 324, 301-313.

Johnson, R. E., J. G. Luhmann, R. L. Tokar, M. Bouhram, J. J. Berthelier, E. C. Sittler, J. F. Cooper, T. W. Hill, H. T. Smith, M. Michael, M. Liu, F. J. Crary and D. T.

Young (2006) Production, ionization and redistribution of $O_2$ in Saturn's ring atmosphere. *Icarus,* 180, 393-402.

Jones, G. H., N. Krupp, H. Krüger, E. Roussos, W. H. Ip, D. G. Mitchell, S. M. Krimigis, J. Woch, A. Lagg, M. Fränz, M. K. Dougherty, C. S. Arridge and H. J. McAndrews (2006) Formation of Saturn's ring spokes by lightning-induced electron beams. *Geophysical Research Letters,* 33, L21202.

Jones, T. B. (1995) Force interactions between particles. In *Electromechanics of Particles*, 181-217. Cambridge: Cambridge University Press.

Kendall, K., N. M. Alford and J. D. Birchall (1987) A new method for measuring the surface energy of solids. *Nature,* 325, 794-796.

Kimura, H., H. Senshu and K. Wada (2014) Electrostatic lofting of dust aggregates near the terminator of airless bodies and its implication for the formation of exozodiacal disks. *Planetary and Space Science,* 100, 64-72.

Kimura, H., K. Wada, H. Senshu and H. Kobayashi (2015) Cohesion of Amorphous Silica Spheres: Toward a Better Understanding of The Coagulation Growth of Silicate Dust Aggregates. *The Astrophysical Journal,* 812, 67.

Kok, J. F. and N. O. Renno (2006) Enhancement of the emission of mineral dust aerosols by electric forces. *Geophysical Research Letters,* 33, L19S10.

Lebedev, N. N. and I. P. Skalskaya (1962) Force acting on a conducting sphere in field of a parallel plate condenser. *Soviet physics. Technical physics,* 7, 268-270.

Lee, P. (1996) Dust Levitation on Asteroids. *Icarus,* 124, 181-194.

Lemmon, E. W. and S. G. Penoncello (1994) The Surface Tension of Air and Air Component Mixtures. In *Advances in Cryogenic Engineering,* ed. P. Kittel, 1927-1934. Boston, MA: Springer US.

Li, D. et al. (2020) In Situ Investigations of Dust Above the Lunar Terminator at the Chang'E-3 Landing Site in the Mare Imbrium. *Geophysical Research Letters* 47, Issue17, e2020GL089433.

Manka, R. H. (1973) Plasma and Potential at the Lunar Surface. In *Photon and Particle Interactions with Surfaces in Space: Proceedings of the 6th Eslab Symposium,* ed. R. J. L. Grard, 347-361. Dordrecht: Springer Netherlands.

Marouf, E. A., G. L. Tyler, H. A. Zebker, R. A. Simpson and V. R. Eshleman (1983) Particle size distributions in Saturn's rings from voyager 1 radio occultation. *Icarus,* 54, 189-211.

McGhee, C. A., R. G. French, L. Dones, J. N. Cuzzi, H. J. Salo and R. Danos (2005) HST observations of spokes in Saturn's B ring. *Icarus,* 173, 508-521.

Mendis, D., J. R. Hill, H. L. Houpis and E. Whipple (1981) On the electrostatic charging

of the cometary nucleus. *The Astrophysical Journal,* 249, 787-797.

Mitchell, C. J., C. C. Porco, H. L. Dones and J. N. Spitale (2013) The behavior of spokes in Saturn's B ring. *Icarus,* 225, 446-474.

Mitchell, C. J., M. Horányi, O. Havnes and C. C. Porco (2006) Saturn's Spokes: Lost and Found. *Science,* 311, 1587-1589.

Morishima, R., L. Spilker, H. Salo, K. Ohtsuki, N. Altobelli, S. Pilortz (2010) A multilayer model for thermal infrared emission of Saturn's rings II: Albedo, spins, and vertical mixing of ring particles inferred from Cassini CIRS. *Icarus,* 210, 330-345.

Morishima, R., L. Spilker and K. Ohtsuki (2011) A multilayer model for thermal infrared emission of Saturn's rings. III: Thermal inertia inferred from Cassini CIRS. *Icarus,* 215, 107-127.

Morishima, R., L. Spilker, S. Brooks, E. Deau and S. Pilorz (2016) Incomplete cooling down of Saturn's A ring at solar equinox: Implication for seasonal thermal inertia and internal structure of ring particles. *Icarus,* 279, 2-19.

Nakajima, Y. and T. Matsuyama (2002) Electrostatic field and force calculation for a chain of identical dielectric spheres aligned parallel to uniformly applied electric field. *Journal of Electrostatics,* 55, 203-221.

Nicholson, P. D. et al. (2008) A close look at Saturn's rings with Cassini VIMS. *Icarus,* 193, 182-212.

Nordheim, T. A., G. H. Jones, E. Roussos, J. S. Leisner, A. J. Coates, W. S. Kurth, K. K. Khurana, N. Krupp, M. K. Dougherty and J. H. Waite (2014) Detection of a strongly negative surface potential at Saturn's moon Hyperion. *Geophysical Research Letters,* 41, 7011-7018.

Ohtsuki, K. and H. Emori (2000) Local N-Body Simulations for the Distribution and Evolution of Particle Velocities in Planetary Rings. *The Astronomical Journal,* 119, 403.

Ohtsuki, K., H. Kawamura, N. Hirata, H. Daisaka and H. Kimura (2020) Size of the smallest particles in Saturn's rings. *Icarus,* 344, 113346.

Orger, N. C., J. R. C. Alarcon, K. Toyoda and M. Cho (2018) Lunar dust lofting due to surface electric field and charging within Micro-cavities between dust grains above the terminator region. *Advances in Space Research,* 62, 896-911.

Orger, N.C., K. Toyoda, H. Masui, and M. Cho (2021) Experimental investigation on particle size and launch angle distribution of lofted dust particles by electrostatic forces. *Advances in Space Research* 68, 1568-1581.

O'Donoghue, J., L. Moore, J. Connerney, H. Melin, T. S. Stallard, S. Miller and K. H. Baines (2019) Observations of the chemical and thermal response of 'ring rain' on Saturn's ionosphere. *Icarus,* 322, 251-260.

Perko, H. A., J. D. Nelson and W. Z. Sadeh (2001) Surface Cleanliness Effect on Lunar Soil Shear Strength. *Journal of Geotechnical and Geoenvironmental Engineering,* 127, 371-383.

Pieters, C. M. et al. (2009) Character and Spatial Distribution of OH/$H_2$O on the Surface of the Moon Seen by $M^3$ on Chandrayaan-1. *Science,* 326, 568-572.

Pilla, S., J. A. Hamida, K. A. Muttalib and N. S. Sullivan (2008) Thermal hysteresis of the dielectric susceptibility of solid oxygen in the audio frequency range. *Physical Review B,* 77, 224108.

Porco, C. C., J. W. Weiss, D. C. Richardson, L. Dones, T. Quinn and H. Throop (2008) Simulations of the Dynamical and Light-Scattering Behavior of Saturn's Rings and the Derivation of Ring Particle and Disk Properties. *The Astronomical Journal,* 136, 2172.

Porco, C. and G. Danielson (1982) The periodic variation of spokes in Saturn's rings. *The Astronomical Journal,* 87, 826-833.

Poulet, F., D. P. Cruikshank, J. N. Cuzzi, T. L. Roush and R. G. French (2003) Compositions of Saturn's rings A, B, and C from high resolution near-infrared spectroscopic observations. *Astronomy and Astrophysics,* 412, 305-316.

Rennilson, J. J. and D. R. Criswell (1974) Surveyor observations of lunar horizon-glow. *The moon,* 10, 121-142.

Robbins, S. J., G. R. Stewart, M. C. Lewis, J. E. Colwell and M. Sremčević (2010) Estimating the masses of Saturn's A and B rings from high-optical depth N-body simulations and stellar occultations. *Icarus,* 206, 431-445.

Salo, H. (1995) Simulations of Dense Planetary Rings: III. Self-Gravitating Identical Particles. *Icarus,* 117, 287-312.

Scheeres, D. J., C. M. Hartzell, P. Sánchez and M. Swift (2010) Scaling forces to asteroid surfaces: The role of cohesion. *Icarus,* 210, 968-984.

Schmidt, J., K. Ohtsuki, N. Rappaport, H. Salo and F. Spahn (2009) Dynamics of Saturn's dense rings. In *Saturn from Cassini-Huygens,* eds. M. K. Dougherty, L. W. Esposito and S. M. Krimigis, 413-458. Springer.

Schwan, J., X. Wang, H.-W. Hsu, E. Grün and M. Horányi (2017) The charge state of electrostatically transported dust on regolith surfaces. *Geophysical Research Letters,* 44, 3059-3065.

Senshu, H. et al. (2015) Photoelectric dust levitation around airless bodies revised using realistic photoelectron velocity distributions. *Planetary and Space Science* 116, 18-29.

Smith, B. A. et al. (1981) Encounter with Saturn: Voyager 1 imaging science results. *Science,* 212, 163-191.

Sow, M., R. Widenor, A. R. Akande, K. S. Robinson, R. M. Sankaran and D. J. Lacks (2013) The role of humidity on the lift-off of particles in electric fields. *Journal of the Brazilian Chemical Society,* 24, 273-279.

Spilker, L., C. Ferrari and R. Morishima (2013) Saturn's ring temperatures at equinox. *Icarus,* 226, 316-322.

Spitzer Jr, L. (2008) *Physical processes in the interstellar medium.* John Wiley and Sons.

Steinpilz, T., J. Teiser, and G. Wurm (2019) Sticking Properties of Silicates in Planetesimal Formation Revisited. *The Astrophysical Journal* 874, 60.

Thornton, C. and Z. Ning (1998) A theoretical model for the stick/bounce behaviour of adhesive, elastic-plastic spheres. *Powder Technology,* 99, 154-162.

Tokar, R. L. et al. (2005) Cassini observations of the thermal plasma in the vicinity of Saturn's main rings and the F and G rings. *Geophysical Research Letters,* 32, L14S04.

Tseng, W. L., W. H. Ip, R. E. Johnson, T. A. Cassidy and M. K. Elrod (2010) The structure and time variability of the ring atmosphere and ionosphere. *Icarus,* 206, 382-389.

van den Bosch, H. F. M., K. J. Ptasinski and P. J. A. M. Kerkhof (1995) Two conducting spheres in a parallel electric field. *Journal of Applied Physics,* 78, 6345-6352.

von Holstein-Rathlou, C., J. P. Merrison, C. F. Brædstrup and P. Nørnberg (2012) The effects of electric fields on wind driven particulate detachment. *Icarus,* 220, 1-5.

Vigil, G., Z. Xu, S. Steinberg and J. Israelachvili (1994) Interactions of Silica Surfaces. *Journal of Colloid and Interface Science,* 165, 367-385.

Wang, X., J. Schwan, H. W. Hsu, E. Grün and M. Horányi (2016) Dust charging and transport on airless planetary bodies. *Geophysical Research Letters,* 43, 6103-6110.

Wang, X., M. Horányi, Z. Sternovsky, S. Robertson and G. E. Morfill (2007) A laboratory model of the lunar surface potential near boundaries between sunlit and shadowed regions. *Geophysical Research Letters,* 34, L16104.

Xie, L., X. Zhang, L. Li, B. Zhou, Y. Zhang, Q. Yan, Y. Feng, D. Guo, and S. Yu (2020) Lunar Dust Fountain Observed Near Twilight Craters. *Geophysical Research Letters* 47, e2020GL089593.

Young, D. T. et al. (2005) Composition and Dynamics of Plasma in Saturn's Magnetosphere. *Science,* 307, 1262-1266.

Zebker, H. A., E. A. Marouf and G. L. Tyler (1985) Saturn's rings: Particle size distributions for thin layer models. *Icarus,* 64, 531-548.